\documentclass[prl,twocolumn,superscriptaddress,notitlepage]{revtex4-2}

\usepackage{graphicx}
\usepackage{amsmath,amsfonts,amsthm,amssymb} 
\usepackage{times}
\usepackage{bm}
\usepackage{upgreek}
\usepackage[svgnames]{xcolor}
\usepackage{hyperref}
\hypersetup{
    colorlinks=true,
    linkcolor=RoyalBlue,
    citecolor=RoyalBlue,
    filecolor=black,      
    urlcolor=black,
    pdftitle={Overleaf Example},
    pdfpagemode=FullScreen,
    }

\newcommand{\mat}[1]{\bar{\bar{#1}}} 
\newcommand{\matb}[1]{\bar{\bar{\mathbf{#1}}}} 
\newcommand{\bra}[1]{\langle #1|} 
\newcommand{\ket}[1]{|#1\rangle}

\begin{document}

\title{Selective Radiance in Super-Wavelength Atomic Arrays}

\author{Charlie-Ray Mann}
\affiliation{ICFO-Institut de Ciencies Fotoniques, The Barcelona Institute of Science and Technology, 08860 Castelldefels (Barcelona), Spain.}

\author{Francesco Andreoli}
\affiliation{ICFO-Institut de Ciencies Fotoniques, The Barcelona Institute of Science and Technology, 08860 Castelldefels (Barcelona), Spain.}

\author{Vladimir Protsenko}
\affiliation{Jo\v zef Stefan Institute, SI-1000 Ljubljana, Slovenia.}

\author{Zala Lenar\v ci\v c}
\affiliation{Jo\v zef Stefan Institute, SI-1000 Ljubljana, Slovenia.}

\author{Darrick E. Chang}
\affiliation{ICFO-Institut de Ciencies Fotoniques, The Barcelona Institute of Science and Technology, 08860 Castelldefels (Barcelona), Spain.}
\affiliation{ICREA-Instituci\'o Catalana de Recerca i Estudis Avan\c{c}ats, 08015 Barcelona, Spain.}

\begin{abstract}
\noindent A novel way to create efficient atom-light interfaces is to engineer collective atomic states that selectively radiate into a target optical mode by suppressing emission into undesired modes through destructive interference. While it is generally assumed that this approach requires dense atomic arrays with sub-wavelength lattice constants, here we show that selective radiance can also be achieved in arrays with super-wavelength spacing. By stacking multiple two-dimensional arrays we find super-wavelength mirror configurations where one can eliminate emission into unwanted diffraction orders while enhancing emission into the desired specular mode, leading to near-perfect reflection of weak resonant light. These super-wavelength arrays can also be functionalized into efficient quantum memories, with error probabilities on the order of $\sim 1\%$ for a trilayer with only around $\sim100$ atoms per layer. Relaxing the previous constraint of sub-wavelength spacing could potentially ease the technical requirements for realizing efficient atom-light interfaces, such as enabling the use of tweezer arrays.

\end{abstract}

\maketitle

%===========================================================%
%--------------------- INTRODUCTION ------------------------%
%===========================================================%

\noindent \textbf{Introduction.---} Creating an efficient atom-light interface is challenging because atoms are inherently dissipative objects; they can absorb photons from a desired optical mode and scatter them into other inaccessible modes, resulting in a loss of quantum information. Rather than relying on the collective enhancement in disordered ensembles \cite{Hammerer2010} or the Purcell enhancement in structured photonic environments \cite{Chang2018} to improve efficiencies, a new generation of atom-light interfaces have been proposed based on \emph{selective radiance} \cite{Asenjo2017}. The central idea is to exploit wave interference as a new resource in carefully engineered configurations of atoms in order to suppress spontaneous emission into undesired modes. 

A striking example occurs in ordered two-dimensional (2D) atomic arrays with sub-wavelength lattice constants. These arrays can act as a near-perfect mirror for weak resonant light \cite{Bettles2016,Shahmoon2017} and be functionlized into an efficient quantum memory \cite{Manzoni2018} and deterministic photon-photon gate \cite{Moreno2021}. The selective radiance leads to a substantial reduction in errors compared to what is predicted for disordered ensembles \cite{Gorshkov2007,Thompson2017}, where spontaneous emission is treated as an independent process. In particular, errors at the $\sim 1\%$ level should be feasible using sub-wavelength arrays, a goal that has proven elusive for conventional atom-light interfaces \cite{Vernaz2018,Wang2019,Tiarks2019,Stolz2022}.

To realize 2D sub-wavelength arrays, current experiments utilize ultracold atoms in a Mott insulator phase of an optical lattice \cite{Rui2020,Srakaew2023}. A tantilizing alternative would be to employ optical tweezers \cite{Barredo2016,Barredo2018,Scholl2021,Endres2016,Bernien2017,Ebadi2021} as one can assemble defect-free arrays in almost any desired geometry, and they offer significantly shorter experimental cycle times which would lead to faster repetition rates for quantum optics operations. However, unfortunately, the diffraction-limited focus of the individual tweezers presents a serious challenge to achieving sub-wavelength lattice constants. To that end, it is important to establish whether we can relax this constraint and find selectively radiant configurations with \emph{super-wavelength} spacing, which is more compatible with current tweezer technology.

\begin{figure}
\centering
\includegraphics[width=0.46\textwidth]{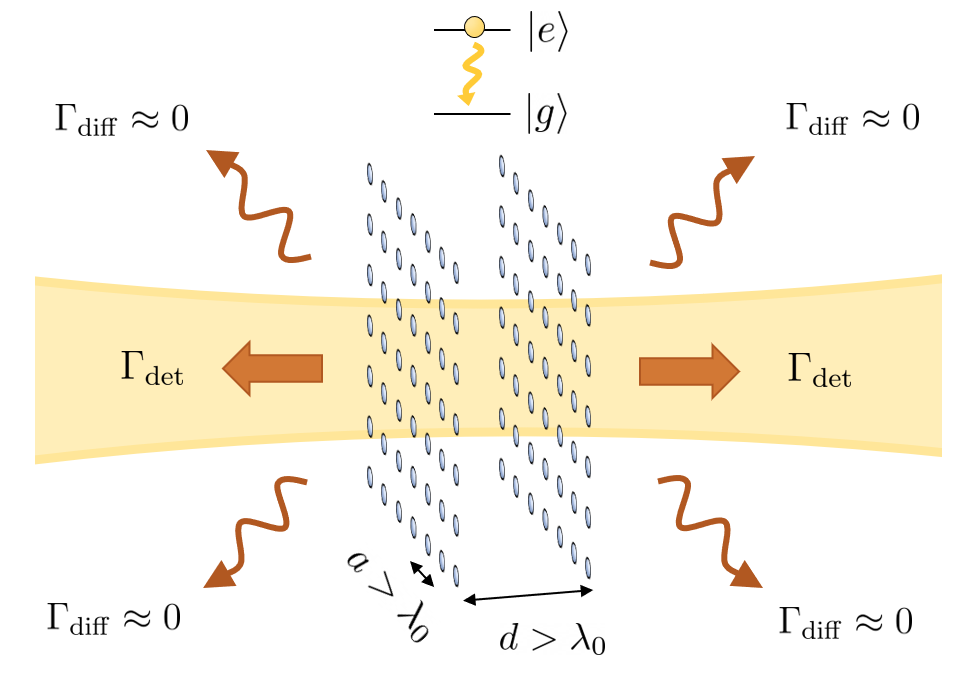}
\caption{Schematic of a super-wavelength mirror composed of a bilayer atomic array with lattice constant $a>\lambda_0$ and layer spacing $d>\lambda_0$. The array can selectively radiate into the desired specular mode at a rate $\Gamma_{\rm{det}}=M\Gamma_{00}$ which is enhanced by the number of layers $M$ ($\Gamma_{00}$ being the rate of a monolayer), while the emission into unwanted diffraction orders is eliminated through inter-layer destructive interference $\Gamma_{\rm{diff}}=0$.}
\label{fig:Schematic_Bilayer}
\end{figure}

While it is well known that a 2D super-wavelength array is a poor light-matter interface, due to photons being scattered into multiple diffraction orders, here we demonstrate that one can restore the selective radiance by stacking multiple 2D layers. Using an idealized model, we find a range of super-wavelength mirror configurations which support collective atomic states that selectively radiate into the desired specular at an enhanced rate, while the emission into all unwanted diffraction orders in eliminated through inter-layer destructive interference (see Fig.~\ref{fig:Schematic_Bilayer}). Guided by this intuition, we show that a finite super-wavelength array can almost perfectly reflect a weak classical beam on resonance, and be functionalized into into an efficient quantum memory for single photons. For example, one can achieve errors on the order of $\sim1\%$ for a trilayer with only around $\sim100$ atoms per layer.

%------------ FIGURE 1 -------------% 
%------------ FIGURE 1 -------------%

%===========================================================%
%------------------ SPIN MODEL FORMALISM -------------------%
%===========================================================%

\textbf{Spin model formalism.---} We consider an ensemble of $N$ identical atoms trapped at fixed positions $\mathbf{r}_i$. Initially, we assume a minimal two-level model where the ground $\ket{g_{i}}$ and excited $\ket{e_{i}}$ states support a circularly-polarized transition with frequency $\omega_{0}$ and dipole matrix element $\bm{\wp}=|\bm{\wp}|\mathbf{e}_{\wp}$, where $\mathbf{e}_{\wp}$ is the unit polarization vector. The ensemble is then driven by a coherent input field with a (near-resonant) central frequency $\omega_{\rm p}$ and amplitude $\mathbf{E}_{\rm in}^+(\mathbf{r})=\langle\hat{\mathbf{E}}_{\rm in}^+(\mathbf{r})\rangle$ whose polarization matches the atomic transition. 

Within the quantum jump formalism and the Born-Markov approximation \cite{Meystre2007}, the atomic state $|\psi\rangle$ evolves deterministically $\mathrm{i}\hbar \partial_t |\psi\rangle = \hat{H}_{\mathrm{eff}}|\psi\rangle$ under the non-Hermitian Hamiltonian $\hat{H}_{\mathrm{eff}}=\hat{H}_{\mathrm{in}}+\hat{H}_{\mathrm{dip}}$ \cite{Asenjo2017}. This should then be supplemented with stochastically applied quantum jumps, but these will not be important for the observables of interest here \cite{Manzoni2018,Moreno2021}. In the rotating frame  the input Hamiltonian reads
%----------------------
\begin{equation}
       \hat{H}_{\mathrm{in}}=-\hbar \sum_{i} \Delta\hat{\sigma}_{ee}^{i}-\hbar\sum_{i}(\Omega_{i}\hat{\sigma}_{eg}^{i}+\mathrm{H.c.})\,,
      \label{eq:Input_Hamiltonian}
\end{equation}
%----------------------
where $\hat{\sigma}_{\mu\nu}^i=\ket{\mu_i}\bra{ \nu_i}$ are the atomic operators with $\{\mu,\nu\}\in\{g,e\}$, $\Delta =\omega_{\rm p}-\omega_{0}$ is the atom-probe detuning, and $\Omega_{i} = \bm{\wp}^*\cdot \mathbf{E}^+_{\rm in}(\mathbf{r}_{i})/\hbar$ are the Rabi frequencies. Furthermore, the dipole-dipole Hamiltonian is \cite{Asenjo2017}
%----------------------
\begin{equation}
       \hat{H}_{\mathrm{dip}}= -\hbar \Gamma_0\sum_{ij}G_{ij}\hat{\sigma}_{eg}^{i}\hat{\sigma}_{ge}^{j}\,,
      \label{eq:Dipole_Hamiltonian}
\end{equation}
%----------------------
where $G_{ij}=\mathbf{e}_{\wp}^*\cdot \matb{G}(\mathbf{r}_{i}-\mathbf{r}_{j},\omega_{0})\cdot\mathbf{e}_{\wp}$ is the projected component of the (dimensionless) dyadic Green function \cite{Novotny2012}
%----------------------
\begin{equation}
\begin{split}
    \matb{G}(\mathbf{r},\omega_{0})=\frac{3\mathrm{e}^{\mathrm{i}k_0r}}{4k_0^3 r^3}\big[\big(k_0^2r^2+\mathrm{i}k_0r-1\big)\mat{\mathbf{I}}\\
    -\big(k_0^2r^2+3\mathrm{i}k_0r-3\big)\hat{\mathbf{r}}\otimes\hat{\mathbf{r}}\big]\,,
    \label{eq:Greens_Function}
\end{split}
\end{equation}
%----------------------
with $k_0=\omega_{0}/c$. This Hamiltonian describes the effective coherent ($\operatorname{Re}\{G_{ij}\}$) and dissipative ($\operatorname{Im}\{G_{ij}\}$) interactions between the atoms which are mediated by the free-space photons. The coherent part of the self interaction, associated with the Lamb shift, is divergent and assumed to be absorbed into the definition of $\omega_0$, while the dissipative part is finite $\operatorname{Im}\{G_{ii}\}=1/2$ and $\Gamma_0=k_0^3|\bm{\wp}|^2/3\pi\hbar \varepsilon_0 $ is the free-space spontaneous emission rate of an isolated atom.

Once the dynamics of the atoms have been solved for, one can reconstruct the field correlations in a target detection mode $\bm{\mathcal{E}}_{\rm det}(\mathbf r)$ which can be efficiently captured in a given optical setup (e.g. a Gaussian). The associated operator is \cite{Mann2023}
%----------------------
\begin{equation}
       \hat{a}_{\mathrm{det}}= \hat{a}_{\mathrm{det,in}}+\mathrm{i}\sqrt{\frac{3\Gamma_0}{8\pi }}\sum_i \bm{\mathcal{E}}_{\mathrm{det}}^*(\mathbf{r}_i)\cdot\mathbf{e}_{\wp} \,\hat{\sigma}_{ge}^{i}\,,
      \label{eq:Detection_Mode}
\end{equation}
%----------------------
where we choose the normalization so that $\langle \hat{a}_{\rm det}^\dagger \hat{a}_{\rm det}\rangle$ corresponds to the rate of outgoing photons emitted into the detection mode. The first and second term in Eq.\,\eqref{eq:Detection_Mode} are the contributions from the input and scattered fields, respectively.

We consider a driving amplitude $\Omega_i$ that is weak enough so that the dynamics are effectively restricted to the single-excitation manifold and the low saturation limit. If we take the detection mode to be the same as the input mode, but propagating in the opposite direction, then the reflection coefficient is $r=-(3\mathrm{i}/8\pi)\vec{\mathcal{E}}\cdot \mat{\Lambda}\cdot\vec{\mathcal{E}}$ and the reflectance is $R=|r|^2$ \cite{Mann2023}. Here, $\vec{\mathcal{E}} =(\mathcal{E}_1,\mathcal{E}_2,\dots,\mathcal{E}_N)^T$ with $\mathcal{E}_i=\mathbf{e}_{\wp} \cdot\bm{\mathcal{E}}_{\mathrm{det}}^*(\mathbf{r}_i)$, and $\mat{\Lambda}$ is the linear response matrix whose elements are given by $(\mat{\Lambda}^{-1})_{ij}=(\Delta/\Gamma_0)\delta_{ij}+G_{ij}$.

%------------ FIGURE 2 -------------% 
\begin{figure*}[t]
\centering
\includegraphics[width=0.92\textwidth]{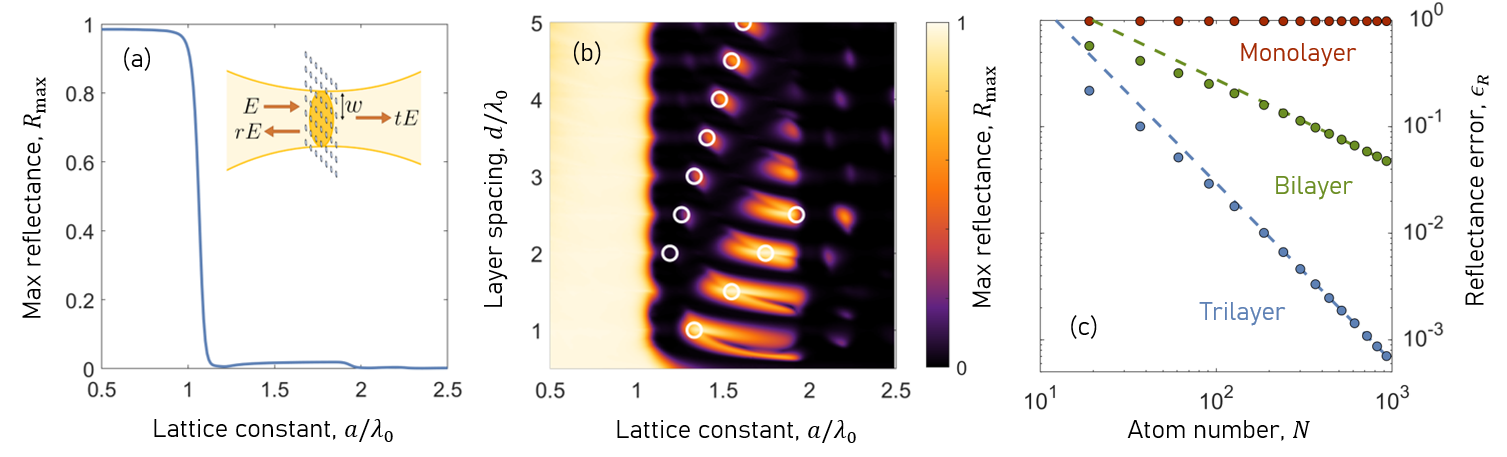}
\caption{(a) Maximum specular reflectance $R_{\rm max}$ of a Gaussian beam as a function of lattice constant for a triangular monolayer ($N=127$, $w=\sqrt{N}a/3$). (b) Maximum reflectance as a function of lattice constant and layer spacing for a trilayer. The peaks are predicted well by the idealized model (white circles). (c) Scaling of the minimum reflectance error $\epsilon_R=1-R_{\rm max}$ as function of $N$ for a monolayer (red dots), bilayer (green dots) and trilayer (blue dots), where we have optimized the set $\{a,d,w\}$ near $(a,d) \approx (1.6\lambda_0,1.5\lambda_0)$. The dotted lines indicate the scaling $\epsilon_R \approx 10.6N^{-0.8}$ and $\epsilon_R \approx 66.6N^{-1.7}$ for the bilayer and trilayer, respectively. }
\label{fig:Resonant_Reflectance}
\end{figure*}
%------------ FIGURE 2 -------------%

%===========================================================%
%----------------- SUB-WAVELENGTH MIRROR -------------------%
%===========================================================%

\textbf{Sub-wavelength mirror.---} It is insightful to first review how a sub-wavelength atomic array can act as a perfect mirror \cite{Bettles2016,Shahmoon2017}, and why this breaks down for super-wavelength spacing. For concreteness, we initially consider a square lattice of atoms located at periodic positions $\mathbf{r}_i$ in the $xy$ plane with lattice constant $a$, although the same arguments apply to any Bravais lattice. In the idealized limit $N \to \infty$ the discrete translational symmetry dictates that the single-excitation atomic eigenstates of Eq.\,\eqref{eq:Dipole_Hamiltonian} are $|\mathbf{q}\rangle = \hat{S}_{\mathbf{q}}^\dagger|g\rangle^{\otimes N}$, where $\hat{S}_{\mathbf{q}}^\dagger=N^{-1/2}\sum_i \mathrm{e}^{\mathrm{i}\mathbf{q}\cdot\mathbf{r}_i}\hat{\sigma}_{eg}^i$ creates a collective spin wave with Bloch wavevector $\mathbf{q}\in \rm{BZ}$. Although the array supports $N$ such eigenstates, a plane wave at normal incidence can only couple to the spin wave with $\mathbf{q}=\mathbf{0}$. However, the in-plane momentum is only conserved modulo a reciprocal lattice vector $\mathbf{g}_{mn}=(2\pi/a)(m,n)$, which we index by $m,n\in \mathbb{Z}$. As a result, the spin wave can scatter photons into a discrete set of propagating diffraction orders $(mn)$ in addition to the specular detection mode $(00)$.

Due to the excitation of a single eigenstate, the reflection coefficient has a simple Lorentzian form \cite{Mann2023}
%----------------------
\begin{equation}
       r=\frac{\mathrm{i}\Gamma_{\mathrm{det}}/2}{-\Delta+J-\mathrm{i}(\Gamma_{\mathrm{det}}+\Gamma_{\mathrm{diff}})/2} \,,
      \label{eq:Array_Reflection}
\end{equation}
%----------------------
where $J=-\Gamma_0\operatorname{Re}\{\sum_{i\neq j}G_{{ij}}\}$ is the collective frequency shift of the spin wave. The corresponding collective decay rate $\Gamma=2\Gamma_0\operatorname{Im}\{\sum_{i}G_{{ij}}\}$ can be expressed as a sum over the different emission channels $\Gamma=\sum_{mn}\Gamma_{mn}$, where
%----------------------
\begin{equation}
      \Gamma_{mn}=\Gamma_{00}\frac{k_0^2+k_{mn}^2}{2k_0 k_{mn}} \Theta (k_0-|\mathbf{g}_{mn}|)\,.
      \label{eq:Decay_Rates}
\end{equation}
%----------------------
Here, $\Gamma_{00}=3\pi \Gamma_0/k_0^2\mathcal{A}$ is the decay rate into the specular detection mode ($\Gamma_{\mathrm{det}}=\Gamma_{00}$), where $\mathcal{A}$ is the area of the unit cell, while $\Gamma_{\mathrm{diff}}=\sum_{mn\neq 00} \Gamma_{mn}$ is the decay rate into all the propagating diffraction orders (see Fig.~\ref{fig:Schematic_Bilayer}). We have introduced the Heaviside step function $\Theta (x)$ in Eq.\,\eqref{eq:Decay_Rates} because $k_{mn} =(k_0^2-|\mathbf{g}_{mn}|^2)^{1/2}$ becomes imaginary for $|\mathbf{g}_{mn}|>k_0$; these correspond to evanescent diffraction orders which do not represent possible emission channels.

The maximum reflectance on resonance ($\Delta=J$) is determined by the branching ratio of the emission, $R_{\rm max}=[\Gamma_{\mathrm{det}}/(\Gamma_{\mathrm{det}}+\Gamma_{\mathrm{diff}})]^2$. If the lattice constant is sub-wavelength $a<\lambda_0$ then all of the diffraction orders become evanescent ($\Gamma_{\mathrm{diff}}=0$), leading to perfect reflection on resonance $R_{\rm max}=1$. For super-wavelength lattice constants $a>\lambda_0$ there is a dramatic reduction in reflectance because multiple diffraction orders open up with $\Gamma_{\mathrm{det}}/\Gamma_{\mathrm{diff}}\ll 1$, thereby losing the selective radiance of the sub-wavelength array.

%===========================================================%
%----------------- SUPER-WAVELENGTH MIRROR -----------------%
%===========================================================%

\textbf{Super-wavelength mirror.---} To restore the selective radiance we now consider $M$ super-wavelength layers, each containing $N$ atoms located at $\mathbf{r}_{i\alpha}=\mathbf{r}_{i}+\mathbf{d}_\alpha$. Here, $\mathbf{r}_{i}$ represent the 2D Bravais lattice vectors and $\mathbf{d}_\alpha=(\bm{\uprho}_\alpha,z_\alpha)$ is the shift of each layer indexed by $\alpha$. To gain some intuition, we again consider the idealized limit $N\to \infty$ and a plane wave input at normal incidence. The relevant single-excitation manifold is now spanned by  $\ket{\alpha}=\hat{S}_{\alpha}^\dagger \ket{g}^{\otimes NM}$, where $\hat{S}_{\alpha}^\dagger=N^{-1/2}\sum_i \hat{\sigma}_{eg}^{i\alpha}$ creates a collective spin wave with $\mathbf{q}=\mathbf{0}$ in layer $\alpha$. Within this manifold the dynamics are effectively described by the Hamiltonian
%----------------------
\begin{equation}
       \hat{H}_{\mathrm{dip}}= -\hbar \Gamma_{00}\sum_{\alpha\beta}\mathcal{G}_{\alpha\beta}\hat{S}_{\alpha}^{\dagger}\hat{S}_{\beta}\,,
      \label{eq:Hamiltonian_Multilayer}
\end{equation}
%----------------------
where the inter-layer matrix elements are
%----------------------
\begin{equation}
        \mathcal{G}_{\alpha\beta} =\frac{\mathrm{i}}{2} \sum_{mn} \frac{\Gamma_{mn}}{\Gamma_{00}}\mathrm{e}^{\mathrm{i}k_{mn}|z_\alpha-z_\beta|} \mathrm{e}^{\mathrm{i}\mathbf{g}_{mn}\cdot (\bm{\uprho}_\alpha-\bm{\uprho}_\beta)}\,.
      \label{eq:Interlayer_Matrix_Elements}
\end{equation}
%----------------------
For simplicity, we have neglected the coherent interactions mediated by the evanescent diffraction orders, and we have also absorbed $J$ into the detuning term.

In general, the input field will couple to all the $M$ collective atomic eigenstates of Eq.\,\eqref{eq:Hamiltonian_Multilayer}, giving rise to complex interference effects. To recover a single-state response we search for the analog of the ``mirror'' configuration in waveguide QED \cite{Chang2012,Mirhosseini2019}, where the coherent inter-layer interactions vanish $\operatorname{Re}\{\mathcal{G}_{\alpha\beta}\}=0$ and the atomic eigenstates are energetically degenerate. With this goal it is favorable to preserve the 2D point symmetry group of the underlying Bravais lattice. Then, if the interactions are dissipative for some $\mathbf{g}_{mn}$ they are also dissipative for all those with the same magnitude $|\mathbf{g}_{mn}|$. This condition is satisfied by stacking parallel layers ($\bm{\uprho}_\alpha=0$), but there also exist other high-symmetry configurations (e.g. square lattices shifted by half a unit cell).

To ensure that the inter-layer interactions mediated by the specular mode are always dissipative, we can fix the layer spacing $d=\ell\lambda_0/2$ with $\ell \in \mathbb{N}$, which means that we only have one parameter to tune -- the lattice constant. We thus consider the simplest case where there is only one set of diffraction orders; for the square lattice this restricts us to the range $a/\lambda_0\in (1,1.41)$, while for the triangular lattice the range is much larger $a/\lambda_0\in (1.15,2)$. From Eq.\,\eqref{eq:Interlayer_Matrix_Elements} one can show that the inter-layer interactions become purely dissipative when
%----------------------
\begin{equation}
        Q=\ell\Big(1+\sqrt{1-|\mathbf{g}_{mn}|^2/k_0^2}\Big)
      \label{eq:Critical_Condition}
\end{equation}
%----------------------
is equal to an integer $Q\in\mathbb{N}$. From this set of critical lattice constants we can distinguish two qualitatively different configurations. 

If $Q\in\mathbb{N}_\mathrm{even}$ then the diffraction orders mediate dissipative interactions with the \emph{same} sign as the specular mode \cite{Mann2023}. Consequently, $\mathcal{G}_{\alpha\beta}$ becomes a rank-1 matrix where there is only one bright eigenstate $\ket{\mathrm{B}}=M^{-1/2}\sum_\alpha(-1)^{\ell\alpha} \ket{\alpha}$ with decay rate $M \Gamma$, while the other $M-1$ eigenstates are completely dark. This bright state is not selectively radiant because the emission into all channels is collectively enhanced, and thus the branching ratio of emission is identical to a monolayer with  $\Gamma_{\mathrm{det}}/\Gamma_{\mathrm{diff}}\ll 1$. Evidently, these are not the desired super-wavelength mirror configurations.

%------------ FIGURE 3 -------------% 
\begin{figure*}[t!]
\centering
\includegraphics[width=0.92\textwidth]{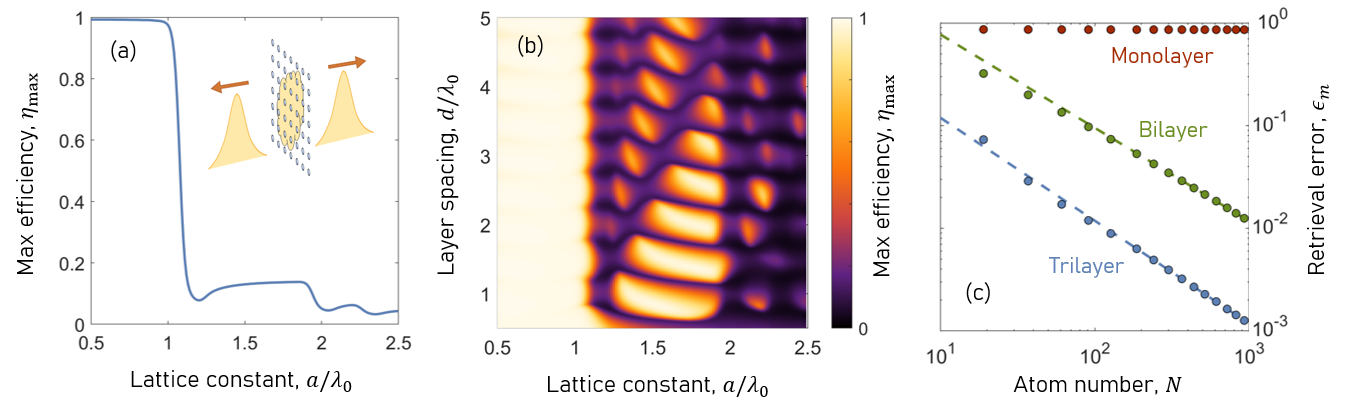}
\caption{(a) Maximum retrieval efficiency $\eta_{\rm max}$ as a function of lattice constant for a triangular monolayer, where we consider a two-way retrieval (see inset) into a Gaussian mode ($N=127$, $w=\sqrt{N}a/3$, $\phi=0$). (b) Maximum retrieval efficiency as a function of lattice constant and layer spacing for a trilayer. (c) Scaling of the minimum retrieval error $\epsilon_m=1-\eta_{\rm max}$ as function of $N$ for a monolayer (red dots), bilayer (green dots) and trilayer (blue dots), where we have optimized the set $\{a,d,w,\phi\}$ near $(a,d) \approx (1.8\lambda_0,1.4\lambda_0)$. The dotted lines indicate the scaling $\epsilon_m \approx 6.3N^{-0.91}$ and $\epsilon_m \approx 1.2N^{-1}$ for the bilayer and trilayer, respectively. }
\label{fig:Retrieval_Efficiency}
\end{figure*}
%------------ FIGURE 3 -------------%

The situation is very different when $Q\in\mathbb{N}_\mathrm{odd}$, because the diffraction orders mediate dissipative interactions with an \emph{alternating} sign with respect to the specular mode \cite{Mann2023}. As a result, $\mathcal{G}_{\alpha\beta}$ becomes a rank-2 matrix with two bright eigenstates and $M-2$ dark eigenstates. For an even number of layers one eigenstate $\ket{\mathrm{B}_1}=M^{-1/2}\sum_\alpha (-1)^{(\ell+1)\alpha} \ket{\alpha}$ has a collectively enhanced decay rate into the diffraction orders $\Gamma_{\mathrm{diff}}=M \sum_{mn\neq00}\Gamma_{mn}$, while the emission into the specular detection mode is completely suppressed $\Gamma_{\mathrm{det}}=0$. In stark contrast, the other eigenstate $\ket{\mathrm{B}_2}=M^{-1/2}\sum_\alpha (-1)^{\ell\alpha} \ket{\alpha}$ has a collectively enhanced decay rate into the specular detection mode $\Gamma_{\mathrm{det}}=M\Gamma_{00}$, while the emission into the diffraction orders is completely suppressed $\Gamma_{\mathrm{diff}}=0$ (see Fig.~\ref{fig:Schematic_Bilayer}). Evidently, the input plane wave can only couple to $\ket{\mathrm{B}_2}$ and this state selectively radiates into the target specular mode --  these are the desired super-wavelength mirror configurations.

Remarkably, even a simple bilayer is sufficient to restore the perfect reflection on resonance, since the high degree of symmetry allows one to shut off emission into all the diffraction orders simultaneously. The physics is slightly more nuanced for an odd number of layers since both bright states contribute to the response, although it stills behaves like a perfect mirror. Moreover, this slight difference between even and odd layers diminishes rapidly with increasing layers \cite{Mann2023}.

%===========================================================%
%------------------- FINITE SIZE EFFECTS -------------------%
%===========================================================%

\textbf{Finite size effects.---} We now consider the realistic case of a finite array and a input Gaussian mode with beam waist $w$ (see inset in Fig.~\ref{fig:Resonant_Reflectance}\textcolor{RoyalBlue}{a}), and study how the errors scale with atom number. For this analysis we focus on triangular lattice configurations since they perform better for a larger range of lattice constants. In Fig.~\ref{fig:Resonant_Reflectance}\textcolor{RoyalBlue}{a} we plot the maximum reflectance as a function of lattice constant for a monolayer with $N=127$ and $w=\sqrt{N}a/3$. For sub-wavelength lattice constants we observe that the reflectance is slightly reduced from unity, which is the result of two fundamental sources of error \cite{Manzoni2018}. First, a small fraction of the Gaussian beam extends beyond the array boundaries which does not interact with the atoms. Second, the Gaussian beam contains a superposition of in-plane wavevectors and can thus couple to spin waves with $\mathbf{q}\neq \mathbf{0}$ which are only quasi-degenerate, thereby losing the ideal single-state response. 

As anticipated, super-wavelength lattice constants exhibit a dramatic reduction in the maximum reflectance because multiple diffraction orders open up. In Fig.~\ref{fig:Resonant_Reflectance}\textcolor{RoyalBlue}{b} we plot the maximum reflectance as a function of lattice constant and layer spacing for a trilayer. We can observe several pockets of high reflectance within the super-wavelength parameter regions, and the peaks are predicted very well by the idealized mirror configurations calculated from Eq.\,\eqref{eq:Critical_Condition}. Note that there are additional sources of error compared to the monolayer case. For example, within the idealized model a perfect mirror exists for any layer separation $d=\ell\lambda_0/2$, but in the realistic case the spatial overlap of the diffracted beams reduces as the layer spacing increases.

In Fig.~\ref{fig:Resonant_Reflectance}\textcolor{RoyalBlue}{c} we show the scaling of the reflectance error $\varepsilon_R=1-R_{\rm max}$ as a function of $N$ for a monolayer, bilayer and trilayer. For each data point we have optimized the set of parameters $\{a,d,w\}$ using a local optimization algorithm, targeting the first super-wavelength mirror configuration near $(a,d) \approx (1.6\lambda_0,1.5\lambda_0)$. As expected for the monolayer, the error plateaus at large values $\epsilon_R \approx 0.98$ because it does not exhibit selective radiance. In contrast, from numerical fitting we find that the reflection error scales as $\epsilon_R \approx 10.6N^{-0.8}$ for the bilayer, while for the trilayer the error scales much faster as $\epsilon_R \approx 66.6N^{-1.7}$. Besides these abstract scalings, with only $N=127$ atoms per layer one can achieve a reflectance of $R_{\rm max}\approx0.79$ with a bilayer and $R_{\rm max}\approx0.98$ with a trilayer -- this represents about a 50-fold increase compared to a monolayer $R_{\rm max}\approx0.02$.

%===========================================================%
%--------------------- QUANTUM MEMORY  ---------------------%
%===========================================================%

\textbf{Quantum memory.---} In the idealized case where the detection mode interacts with a single collective atomic state, it can be established that the resonant reflectance of classical light determines the efficiency of various quantum applications \cite{Solomons2023}. Here, we study the efficiency of an EIT-based quantum memory \cite{Manzoni2018,Gorshkov2007} for the realistic case, where this correspondence does not exactly hold, and study how the errors scale with atom number. The basic idea is to coherently and reversibly map a photonic state to a high-lying, long-lived atomic state $|s_{i}\rangle$, facilitated by a classical control field on the $|s_{i}\rangle \leftrightarrow |e_{i}\rangle$ transition.

It is more convenient to optimize the retrieval of an excitation that is initially stored as a spin wave excitation $|\psi(t=0)\rangle = \sum_{i\alpha} s_{i\alpha}\hat{\sigma}_{sg}^{i\alpha } \ket{g}^{\otimes NM}$, and then the optimal storage process is related via time-reversal symmetry \cite{Gorshkov2007}. The retrieval efficiency is defined as the probability that the photon is emitted into the target detection mode $\eta =\int_0^\infty dt \langle \hat{a}_{\rm det}^\dagger(t) \hat{a}_{\rm det}(t)\rangle $. With the assumption of a spatially uniform control field, one finds $\eta =(3/ 8\pi) \vec{s}\cdot\mat{K}\cdot \vec{s}^{\,*}$
regardless of the temporal profile of the control field \cite{Manzoni2018}. Here, $\vec{s}=(s_{11},\dots,s_{i\alpha},\dots,s_{NM})^T$ is the vector of initial amplitudes and 
%----------------------
\begin{equation}
    \mat{K}=\mathrm{i}\sum_{\xi\xi'}\frac{(\vec{\mathcal{E}}\cdot \vec{v}_{\xi})(\vec{\mathcal{E}}^*\cdot \vec{v}_{\xi'}^{\,*})}{\lambda_{\xi}-\lambda_{\xi'}^*} \vec{v}_{\xi}\otimes\vec{v}_{\xi'}^{\,*}
    \label{eq:KMatrix}
\end{equation}
%----------------------
is a Hermitian matrix, where $\lambda_\xi$ and $\vec{v}_\xi$ are the set of eigenvalues and eigenvectors of the matrix $G_{i\alpha , j\beta}$, which satisfy the relations $\vec{v}_{\xi}\cdot\vec{v}_{\xi'}=\delta_{\xi\xi'}$ and $\sum_\xi \vec{v}_{\xi}\otimes\vec{v}_{\xi}=\mat{I}$. The maximum efficiency $\eta_{\rm max}$ for a given detection mode and set of atomic positions is then given by the maximum eigenvalue of $\mat{K}$, and the corresponding eigenvector gives the optimal initial spin wave excitation \cite{Manzoni2018}.

Since the array will naturally emit outgoing photons in both directions, it is favorable to consider a two-way retrieval scheme  (see inset in Fig.~\ref{fig:Retrieval_Efficiency}\textcolor{RoyalBlue}{a}), where the detection mode is taken to be a superposition of two counter-propagating Gaussian modes with a relative phase $\phi$. In Fig.~\ref{fig:Retrieval_Efficiency}\textcolor{RoyalBlue}{a} we plot the maximum retrieval efficiency as a function of lattice constant for a monolayer with $N=127$, $w=\sqrt{N}a/3$ and $\phi=0$. We see a significant reduction of the retrieval efficiency for super-wavelength lattice constants due to emission into unwanted diffraction orders. In Fig.~\ref{fig:Retrieval_Efficiency}\textcolor{RoyalBlue}{b} we show the efficiency for a trilayer as a function of lattice constant and layer spacing, and we observe many pockets of high efficiency which approximately correlate with the regions of high reflectance in Fig.~\ref{fig:Resonant_Reflectance}\textcolor{RoyalBlue}{b}. 

In Fig.~\ref{fig:Retrieval_Efficiency}\textcolor{RoyalBlue}{c} we show the scaling of the minimum retrieval error $\epsilon_m=1-\eta_{\rm max}$ as a function of $N$ for a monolayer, bilayer and trilayer. For each data point we have locally optimized the set of parameters $\{a,d,w,\phi\}$ targeting the first super-wavelength peak near $(a,d) \approx (1.8\lambda_0,1.4\lambda_0)$. The error for the monolayer plateaus to large values $\epsilon_m \approx 0.86$, but the selective radiance is restored in the bilayer and numerically we find that the error scales as $\epsilon_m \approx 6.3N^{-0.91}$. For the trilayer, the errors are significantly reduced compared to the bilayer, but they scale with a similar power $\epsilon_m \approx 1.2N^{-1}$. While this abstract scaling is similar to what is predicted from Maxwell-Bloch theory for disordered ensembles \cite{Gorshkov2007}, the absolute error for a given atom number is substantially smaller. For the trilayer one can achieve an error of less than $1\%$ with only $N=127$ atoms per layer. To reach such an error with a disordered ensemble would demand a very large optical depth ($\sim 600$) which is technically very challenging to work with, and state-of-the-art errors remain at the $\sim 10\%$ level \cite{Vernaz2018,Wang2019}.

%===========================================================%
%----------------------- CONCLUSIONS -----------------------%
%===========================================================%

\textbf{Conclusion.---} Despite it being a common assumption, we have demonstrated that sub-wavelength spacing is not a fundamental requirement for selective radiance. In fact, one can build efficient light-matter interfaces with super-wavelength arrays which could enable the use of tweezer arrays for quantum optics applications. In principle, one is not limited to periodic configurations with tweezers and the errors can be reduced further by locally optimizing the individual atomic positions. Moreover, it may be fruitful to employ a more sophisticated global optimization algorithm to search for non-intuitive configurations with super-wavelength spacing. Similar to their sub-wavelength counterparts, super-wavelength arrays can be further functionalized with Rydberg interactions to enable an efficient, deterministic photon-photon gate \cite{Moreno2021}.

%===========================================================%
%-------------------- ACKNOWLEDGEMENTS ---------------------%
%===========================================================%

\vspace{10mm}

\textbf{Acknowledgements.---} C.-R.M. acknowledges funding from the Marie Skłodowska-Curie Actions Postdoctoral Fellowship ATOMAG (grant agreement No. 101068503). Z.L. acknowledges  the QuantERA grant QuSiED by MVZI (QuantERA II JTC 2021) and ERC StG 2022 project DrumS, Grant Agreement 101077265. D.E.C acknowledges support from the European Union, under European Research Council grant agreement No 101002107 (NEWSPIN), FET-Open grant agreement No 899275 (DAALI) and EIC Pathfinder Grant No 101115420 (PANDA); the Government of Spain under Severo Ochoa Grant CEX2019-000910-S [MCIN/AEI/10.13039/501100011033]; QuantERA II project QuSiED, co-funded by the European Union Horizon 2020 research and innovation programme (No 101017733) and the Government of Spain (European Union NextGenerationEU/PRTR PCI2022-132945 funded by MCIN/AEI/10.13039/501100011033); Generalitat de Catalunya (CERCA program and AGAUR Project No. 2021 SGR 01442); Fundaci{\'o} Cellex, and Fundaci{\'o} Mir-Puig.

%===========================================================%
%----------------------- REFERENCES ------------------------%
%===========================================================%

\bibliography{Main}

\end{document}